# Light control through a nonlinear lensing effect in a colloid of biosynthesized Gold Nanoparticles

A. Balbuena Ortega[1], E. Brambilia[1],   V. López Gayou[2], R. Delgado Macuil[2], A. Orduña Diaz[2], A. Zamilpa Alvarez[3], A. V. Arzola[1]  and K. Volke-Sepúlveda[1]

[1]Instituto de Física, Universidad Nacional Autónoma de México, Apdo. Postal 20-364, 01000 Cd. de México, México.

[2] IPN -CIBA Tlaxcala, Km. 1.5 Carretera Estatal Tecuexcomac-Tepetitla. Tepetitla de Lardizabal, C.P. 90700, Tlaxcala, México.

[3]CIBIS-IMSS  Calle Rep. Argentina 1,  C.P. 62790, Xochitepec, Morelos, México.

Corresponding author : abalbuena@fisica.unam.mx

Biosynthesis of four samples of colloidal suspensions of gold nanoparticles is achieved using hydroalcoholic extract and three different separated compounds of the plant Bacopa *procumbens*. The nonlinear optical properties of each sample are characterized with the Z-scan technique. In all cases, the Z-scan curves indicate a negative or self-defocusing response, which is mainly attributed to thermal effects. Among the four samples, the hydroalcoholic extract was noted to have the highest nonlinear optical response and was selected to demonstrate the formation of self-collimated beams (SCBs). This kind of beams are obtained when a convergent CW laser, with only few tens of milliwatts of optical power, is introduced into the sample and induces a negative-lens effect that shifts the focal spot forward. As a result, the otherwise highly focused beam propagate with little divergence over lengths of up to 10mm.  Moreover, an SCB is capable of controlling and steering a weak probe beam of a different wavelength, since the probe experiences the lensing induced by the pump. Noteworthy, the response time of the material was found to be less than 0.07s, which makes it a plausible candidate for photonic applications.

## Introduction

Metal nanoparticles are intensely studied nowadays due to their unique optical, electrical and catalytic properties. In order to utilize and optimize their chemical or physical properties, a wide spectrum of research has been focused on controlling their size and shape, which are crucial tuning parameters (1, 2). In recent years, new paths for the synthesis of nanomaterials have been explored through the use of biological materials, like plant extracts (3). This approach offers multiple benefits, such as being eco-friendly and dispensing with the use of toxic chemicals. Furthermore, it allows a controlled synthesis of nanoparticles with well-defined size and shape in a cost-effective way, suitable for large-scale production (4,5). Although biosynthesis of metallic nanoparticles with plants like Cinnamom camphora (6), Nemm (7), Jathopha curcas (8), Jatropha seedcake (9), Bacopa monieri (10), and Bacopa procumbens (11) has been reported, their nonlinear optical properties have been barely explored so far (12-13). There are many interesting applications of metallic nanoparticles though, which rely precisely on exploiting their nonlinear optical properties. For example, the formation of bright optical spatial solitons with propagation distances of several millimeters has been reported using specially tailored metallic nanoparticles of different sizes, shapes and composition, suspended in a liquid, using either CW or pulsed lasers (14,15). While colloidal suspensions, in general, present a much faster response than liquid crystals and photorefractive crystals (16), the required power levels to induce a nonlinear optical response are considerably higher (14-17). However, metallic colloids allow soliton formation with larger propagation distances and less power than in dielectric colloids (18). The nonlinear effects in the latter are due to optical trapping forces (16,17), whereas in the former are mainly attributed to plasmonic resonances.

Although the optical nonlinearity of gold nanoparticles in (14) is shown to be positive or self-focusing up to certain values of the incident power (~150mW), it becomes negative for higher power values. The self-focusing is attributed to the optical gradient force, which attracts the particles with positive polarizability towards the axis of an incident Gaussian beam, whereas the negative response is attributed to thermal effects. Indeed, there are several reports in the literature on the negative thermal nonlinearity of gold nanoparticles (19-21). This does not imply a contradiction, but it is the size and shape of the particles what might come into play to ultimately determine the predominant behaviour.

In this work, we report a thorough experimental investigation of the nonlinear optical properties of four samples of gold nanoparticles synthesized using hydroalcoholic extract and three different compounds of the plant Bacopa procumbens. The obtained samples are characterized by UV-Vis and transmission electron microscopy (TEM). The different compounds of the plant extract were selected in order to modify the size of the nanoparticles, resulting in samples with mean particle diameters between 3 and 5 nm. The Z-scan technique is used for analyzing the behavior of the nonlinear refractive index and the nonlinear absorption. As we used a wavelength sufficiently close to the plasmon resonance and the particles are small enough to neglect the optical forces, the thermal effects are predominant. Finally, we demonstrate the application of these nanocolloids for the generation of self-collimated beams (SCBs) with optical powers of only few tens of milliwatts, which can be used to control a weaker probe beam of a different wavelength. In contrast with previous works, reporting bright soliton formation with powers from hundreds of mW up to some watts in a CW laser (14) or pulses with a very high peak power, of the order of 105 W (15), the mechanism of formation of the SCBs here is a negative-lens effect, which shifts forward the focal position of a convergent input beam.

1. **Biosynthesis of Gold nanoparticles**

A sample of the Bacopa procumbens (Ground or Milled) green plant, collected in San Miguel Regla, Hidalgo, Mexico, was dried and ground. Polar components were extracted using 40g of plant powder and hydroalcoholic solution 50:50 (600 ml) incubated at 76°C, during three periods of 4 hours each in a reflux system (22). Total hydroalcoholic extract (HAE) was obtained after removal of the solvent under vacuum at 50°C. The hydroalcoholic extract was bipartitioned in water/ethyl acetate compounds, by separating the organic fraction from the aqueous solution using different density. In this way, compounds (mixtures) of larger polarity remained in the aqueous fraction (F1) while those of low and medium polarity were obtained in the ethyl acetate fraction (F2)

Tests of synthesis were performed on both fractions, observing that only the ethyl acetate fraction was capable of carrying out the biosynthesis. After that, the separation process of this fraction was done through standard open column chromatography, by introducing F2 to a microporous resin column. The separation of the compounds was performed, in a first instance, in normal phase (70-230 mesh, Merck, Darmstadt, Germany) using a gradient of dichloromethane/methanol for eluting the column with an increase in polarity of 5%.

A total of 49 fractions were obtained and concentrated in a rotatory evaporator under reduced pressure. Fractions were grouped according to compounds similarity and 10 fractions were mixed in this step. Then, they were subjected to a chromatographic fractionation in column of reverse phase (RP-18, 40-63 µm, Merck) with a mobile phase water/acetonitrile in gradient system, obtaining 50 fractions in this second step, that were grouped again according to compounds similarity, until finally obtain three compounds families: Glycosides, Coumarins and Saponins. The compounds obtained from the separation were monitored and identified by thin-layer chromatography (TLC) (23). The activity in the chromatography was monitored to select the family of compounds that allow biosynthesis.

The biosynthesis of gold nanoparticles was prepared using a volume ratio of 7:3 of: gold salt ($HAuCl_4$) and the biological material respectively. The solution of gold salt was prepared to 1mM, while the extracts were prepared to 1mg/ml. The biosynthesis was performed at room temperature for half an hour, obtaining colloidal samples.

## 2. Material Characterization

The thin-layer chromatography plates used to monitor the separation process of the compounds is shown in Fig. 1(a). On the left (I), we can observe the bipartition of the HAE. The aqueous fraction (Fig. 1(a), Lane 1) presents the lowest number of compounds, according to the bands showed in this figure, compared with the organic fraction (Fig. 1(a), Lane 2), which keeps the majority of the compounds founded in the organic fraction.

Part II of Fig. 1(a) shows the thin layer chromatographic summary of the isolated compounds. It can be observed that the organic fraction (Lane A), shows several compounds which are present in the whole extract. Three families of compounds with different retention factor were obtained and identified as: Glycosides (Lane B), Coumarins (Lane C) and Saponins (Lane D). The samples A, B, C and D obtained under these process were used to the biosynthesis.

When the biosynthesis of the compounds was done, all of them showed activity presenting a change in coloration from yellow to ruby red. The monitoring of nanoparticle formation was carried out by UV-Vis spectroscopy, using a Thermo spectrophotometer (Evolution 600). In the UV-Vis curves for the absorbance, Fig 1(b), it is possible observe the peak of the surface plasmon resonance, characteristic of gold nanoparticles. These peaks are slightly shifted for the different samples, within the range from 527.0 to 547.5 nm, due to the nanoparticle size. In the case of sample D, the additional absorption band appearing around 750 nm is characteristic of the Saponins used in the synthesis. The Coumarins used in sample C also exhibit a typical absorption band around 945 nm.

The morphology of the produced gold nanoparticles (AuNPs) was studied using a High Resolution Transmission Electron Microscope (HRTEM, JEM-ARM200CF). The samples for TEM analysis were prepared on carbon coated copper grids. Figure 1(c) shows some typical bright-field TEM images along with the size distribution histograms for the biosynthesized gold nanoparticles with HAE (sample A, top row) and the glycoside compounds (sample D, bottom row). From the histograms, it is seen that the mean size of the gold nanoparticles is around 3 and 4 nm, and they are approximately spherical in shape.

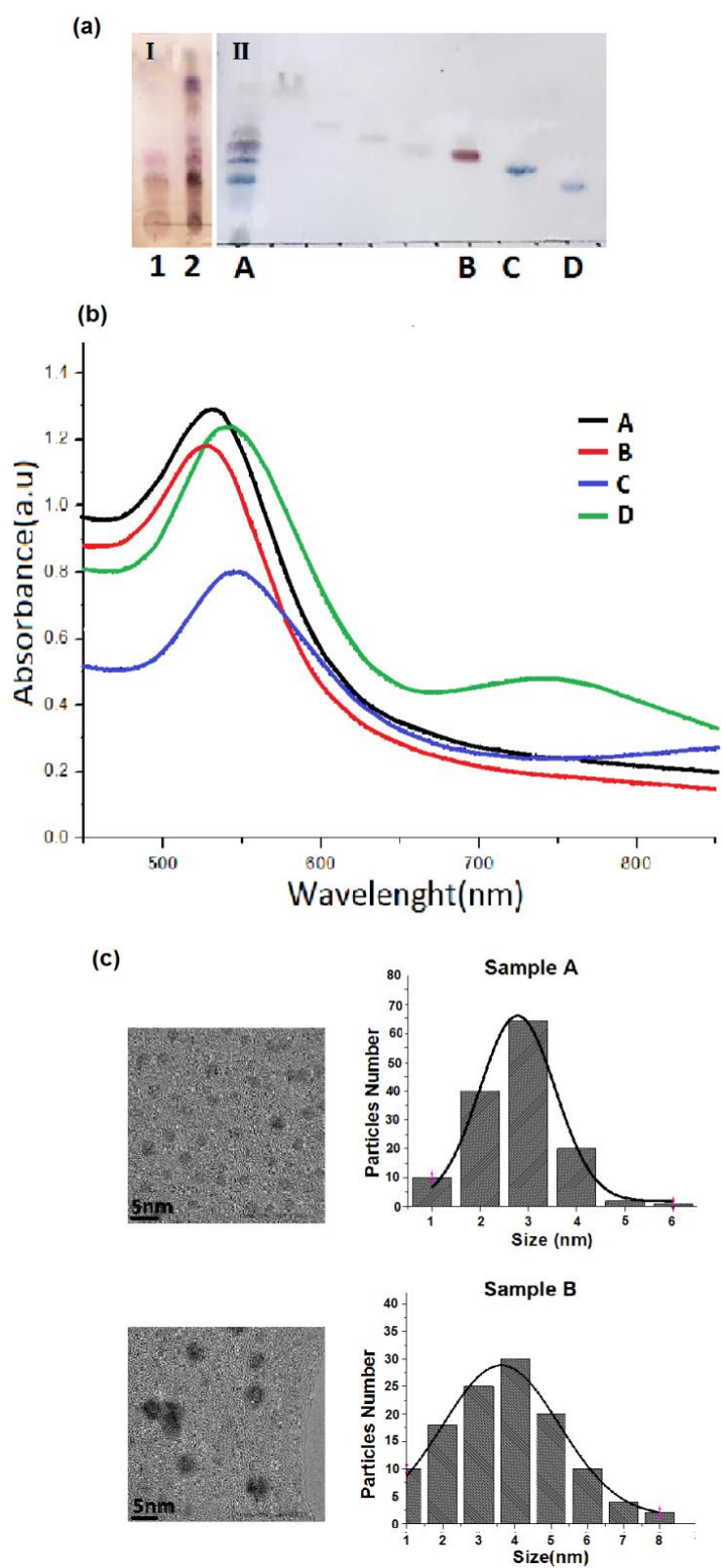

Figure 1. (a) Thin-layer chromatography of: (I) Aqueous Fraction (Lane 1) and Ethanolic Fraction (Lane 2), and (II) isolated compounds (Lanes A, B, C, D). (b) UV-Vis curves for the four obtained samples. (c) HRTEM images and histograms for the biosynthesized gold nanoparticles.

The nonlinear optical properties of the obtained colloids were investigated using the Z-scan technique (24) in the CW regime for a wavelength of 532 nm. This technique provides information about the nonlinear change in the refractive index by means of the so-called closed-aperture detection scheme, while the nonlinear absorption can be attained with the open-aperture detection scheme. In our case, the laser beam was focused onto the sample by a convergent lens with a focal length of 2.54 cm, generating a beam waist of 4.5μm, which leads a Rayleigh range of 120μm. First, the normalized transmittance was measured with a photo-detector through a small aperture as a function of the sample position along the beam axis (z), in order to determine the changes in the nonlinear refractive index. Experimental results for the different samples are shown in Fig. 2 (a) (markers) using an incident power of 60mW, which is the lower threshold to obtain a nonlinear optical response in all the samples. Sample A is clearly identified as the one with the best response, so we selected it for further experiments. A more detailed characterization was performed for this sample at different power values, presented in Fig. 2(b). The maximum change in the refractive index, Δn, has a linear dependence on the incident power, as can be seen in the inset on the right side. The results of the Z-scan in the open-aperture detection scheme showed that there is no nonlinear absorption for any sample at the CW regime, which is consistent with previous reports for gold nanocolloids (19-21).

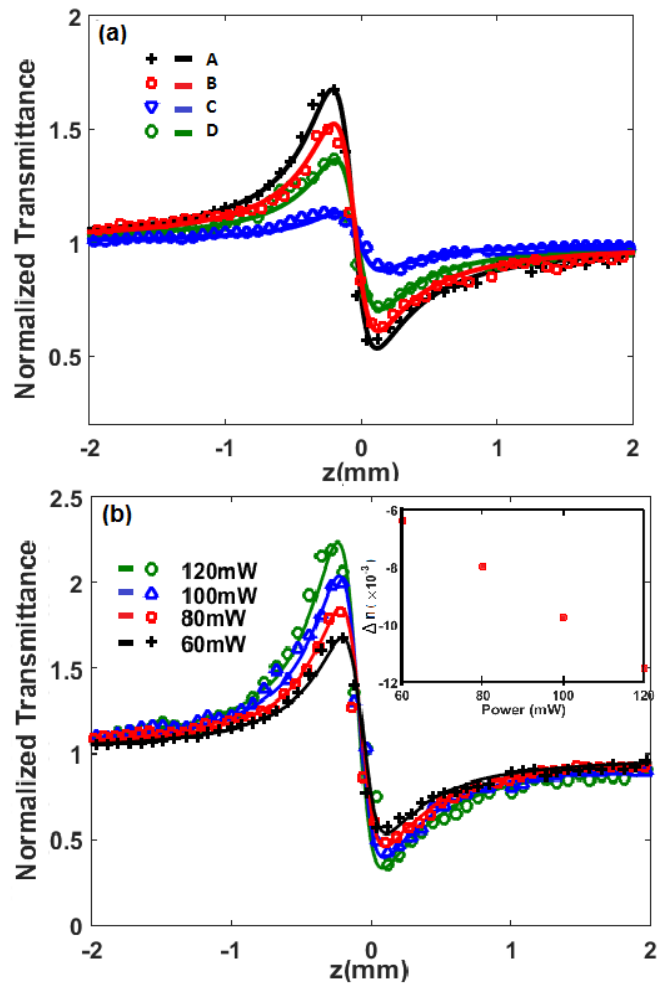

Figure. 2. Z-scan curves for: (a) samples A, B, C and D obtained for an incident power of 60 mW and (b) sample A at different powers, from 60 to 120mW. Markers represent the experimental data, while the solid lines represent the theoretical fitting.

A peak followed by a valley in the Z-scan measurements is indicative of a negative nonlinearity or self-defocusing media (24). The origin of the optical nonlinearity in gold nanoparticle suspensions is threefold (21): there is an ultrafast component of electronic origin; there is a relatively slower component arising from the heating of the metal lattice; and there is a much slower component resulting from the heating of the host medium. Only the third one comes into play in our experiments, since the other two require much higher peak intensities from pulsed lasers and become important at very short times. In our case, the laser light, being strongly absorbed due to the plasmon resonance, gives rise to local heating followed by a heat diffusion process in the medium (20). The lower the thermal conductivity of the host medium the higher the thermal nonlinearity, leading to a nonlocal response, because the spatial temperature profile can differ considerably from the Gaussian profile of the incident light beam.

Accordingly, the model of thermal lens seems to be a good choice for describing the Z-scan signal for gold nanocolloids in the CW regime (20). However, the best fitting results have been obtained with a more general nonlocal model (19). Moreover, closed analytical expressions for the normalized transmittance of nonlinear nonlocal media were presented in (25), which will be used here to fit our experimental results.

Consider a Gaussian beam of wavelength λ, waist radius w0 and Rayleigh range $Z_0$, impinging on a thin nonlinear sample of width L, such that L<< $Z_0$. The output field can be expressed as (26):

$$E_{out}(\rho, z) = E_{in}(\rho, z)\exp[-i\Delta\phi(\rho)] \quad (1)$$

where $\rho=(x^2+y^2)^{1/2}$, z represents the position of the sample along the propagation axis and $\Delta\phi(\rho)$ is the nonlinear phase change, which in the case of a nonlocal medium can be approximated by (27):

$$\Delta\phi(\rho) = \frac{\Delta\Phi_0}{[1+(z/Z_0)]^{m/2}} \exp\left(-\frac{m\rho^2}{w(z)^2}\right) \quad (2)$$

Here, $\Delta\Phi_0$ is the maximum photo-induced phase shift, which takes place on the beam axis, at z=0. The parameter m accounts for the nonlocality of the medium and can take any positive real value. Only for m = 2, the change in the nonlinear phase mimics the transverse intensity distribution of the incident beam, and thus the response of the material is local. Based on this approach, the normalized on-axis transmittance observed at a distance d at the far field (d>>$Z_0$) is found to be (25):

$$T(z, \Delta\Phi_0) = \left|\sum_{n=0}^{N} \frac{1}{n!}\left[\frac{\Delta\Phi_0}{(1+x^2)^{m/2}}\right]^n \frac{i^n(x+i)}{x+i(mn+1)}\right|^2 \quad (3)$$

where $x=z/Z_0$. This expression was used to model the Z-scan measurements. The values of m and $\Delta\Phi_0$ were chosen so as to obtain the best fitting of the experimental data, corresponding to the solid curves shown in Fig. 2. Finally, the nonlinear change in the refractive index is calculated as (24):

$$\Delta n = \frac{\Delta\Phi_0}{kL_{eff}} \quad (4)$$

where $L_{eff}=(1-e^{-\alpha L})/\alpha$, and α represents the linear absorption coefficient.

The properties of the four samples analyzed in this study are summarized in Table 1. We found that the value m=0.3 gave the best fitting for all the samples, meaning that they have the same kind of nonlinear optical response, and since m < 2, we confirm that the response is nonlocal. The value of $\Delta\Phi_0$ is directly related with the magnitude and sign of the change in the nonlinear refractive index. Therefore, samples

A and C exhibit the highest and lowest nonlinear optical response, respectively. If the nonlinearity of the samples is assumed to be Kerr-like, with a dependence of the refractive index on the intensity of the form $n(I)=n_0+ n_2I$, the nonlinear coefficient would be given by $n_2=\Delta n/I_{max}$, where $I_{max} \approx 10^5$ W/cm$^2$, is the maximum intensity at the focal spot. Therefore, in all the samples described in Table 1, $n_2$ would be of the order of $10^{-8}$ cm$^2$/W, which is consistent with previous observations (20).

Table 1. Properties of the different colloidal samples of gold nanoparticles. The nonlinear optical properties (4$^{th}$ and 5$^{th}$ columns) were obtained for 60mW of CW incident optical power.

| Sample | Compound | Nanoparticles Mean Diameter (nm) | $\Delta\Phi_0$ (rad) | $\Delta n$ ( x10$^{-3}$) |
|---|---|---|---|---|
| A | HAE | 2.8±0.8 | -0.72π | -6.4 |
| B | Glycosides | 3.6±1.6 | -0.52π | -4.6 |
| C | Coumarins | 3.5±1.3 | -0.2π | -1.8 |
| D | Saponins | 4.8±1.75 | -0.4π | -3.5 |

## 3. Self-collimated beam: Formation and characterization

As the nonlinear optical response of our metallic nanocolloids is negative, bright spatial solitons cannot be formed. Instead, the self-defocusing effect generates a negative-lens effect within the medium, whose focal length is power-dependent. Therefore, by introducing a convergent beam into the sample, we found that the self-induced negative lens shifts the focal point forward. As a result, the otherwise highly focused beam propagates with little divergence over lengths of up to 10mm, forming what we call a self-collimated beam (SCB). In other words, convergence and self-defocusing can be balanced to some extent, depending on the incident power, to produce a beam that remains approximately collimated over longer distances than in the linear regime. Furthermore, the self-induced lens can act also on a weaker probe beam, controlling its direction and collimation length, as we shall see in the next section.

For these experiments we chose the sample A, which not only exhibited the largest nonlinear refractive index change, but also proved to be the most stable over time. It lasts with negligible changes for almost 90 days, in contrast with other samples that generate agglomerates.

To investigate the formation of SCBs, we used the setup depicted in Fig. 3 in the absence of the probe beam module, enclosed in dashed lines. Namely, a green CW laser (λ=532nm), with an initial diameter of 2.25mm, is focused with a microscope objective (MO) (Olympus Plan N 4X, NA=0.10). As a reference, in the linear regime and propagating in air, the focal spot was measured to be at 1.9 mm from the output of MO, reaching a diameter of ~6.7μm, thus having a Rayleigh range of the order of 256μm.

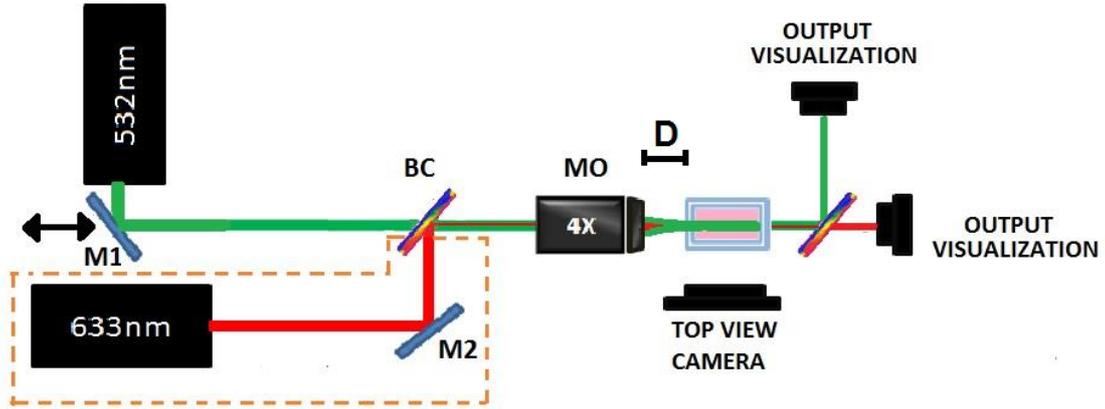

Figure 3. Experimental Set-up: Lasers (532nm, 633nm), mirrors (M1, M2), beam combiner (BC), microscope objective (MO). D represents the distance between the output of the MO and the entrance face of the sample cell. The dashed lines enclose the module for the probe beam experiments.

The length of the sample cell along the z-direction is 10mm, and the beam propagating within sample can be observed and recorded from the top thanks to the scattered light, as shown in Fig. 4(a). The sample is mounted on a linear motorized stage, so it can be moved a distance D from the output of MO along the z-axis (indicated in Fig. 3). The three images in Fig 4(a) illustrate how the beam changes within the sample for different values of D: 15.5, 16.5 and 17.5mm, from top to bottom. This parameter is very relevant because it determines the size (and hence the intensity), as well as the wavefront curvature of the beam impinging on the sample (16). In the linear regime, the position of the focal spot within the sample would be entirely defined by D, whereas in the nonlinear regime it also depends on the incident power, due to the self-induced negative-lens effect.

In a first set of experiments, the beam power measured at the entrance of the sample cell was kept fixed at P=70mW. From each image in Fig. 4(a), a curve for the beam width as a function of $z$ can be obtained. For this purpose, the image is transformed into a numerical matrix and the transverse beam profile is retrieved for each value of the $z$ coordinate. Lorentzian functions give the best fitting to the profiles, and the beam radius is obtained from the fitted parameters, corresponding to the markers in Fig. 4(b). The solid curves represent a fitting of the experimental data using the equation $W_D(z)=W_{CL}(1+(z/CL)^2)^{1/2}$ for the beam radius as a function of z. Here $W_{CL}$ denotes the minimum radius (at the shifted focal plane) and CL represents a collimation length, analogous to the beam waist and the Rayleigh range of a Gaussian beam in a linear medium, respectively. The positions where the beams have their minimum size, $W_{CL}$, are indicated in Fig. 4(b) for the three experimental conditions.

Then, the two parameters identified for optimizing CL are: the distance between the MO and the input face of the sample cell (D) and the input power (P). The optimum D for a given power corresponds to the beam with a maximum CL, as illustrated in Fig. 4(c) for three values of P: 70, 90 and 110mW. It is worth to stress that each point in this plot corresponds to a fitted curve, as those shown in Fig. 4(b). For P=70mW, there is clearly a maximum CL for D≈16.7mm, but for the other two values of P, CL remains approximately constant up to a certain value of D.

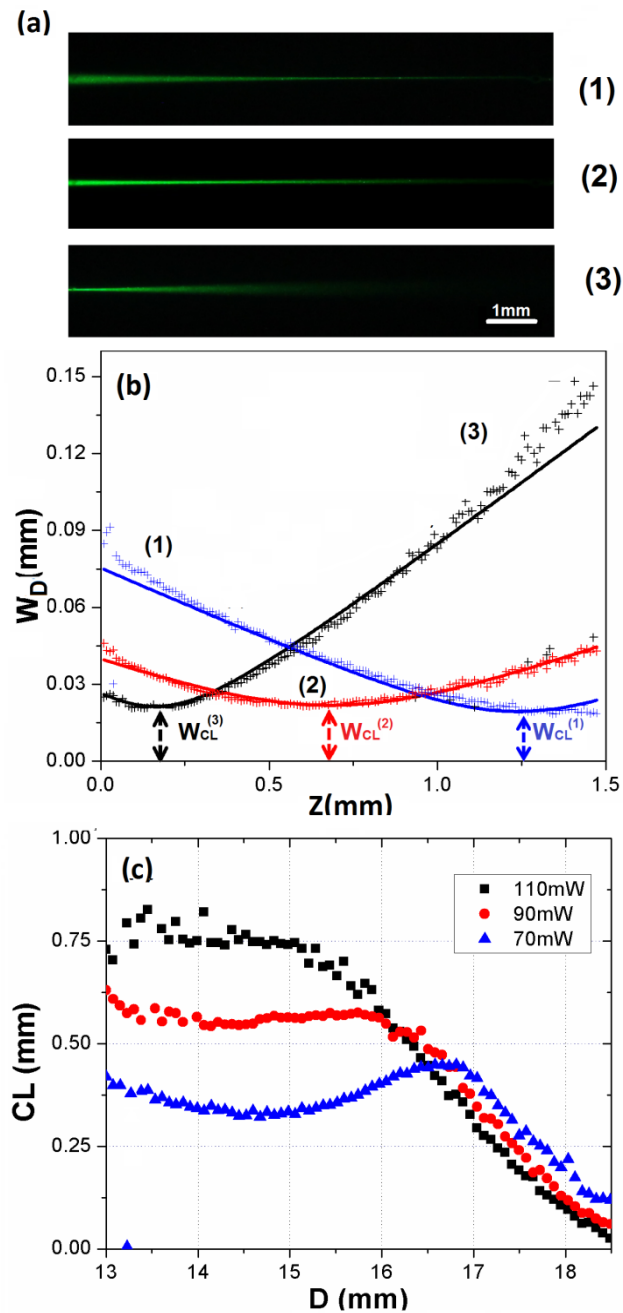

Fig. 4. (a) Top view of the SCB obtained for different values of the distance D: (1) 15.5, (2) 16.5 and (3) 17.5mm. (b) Beam radius inside of the sample as a function of $z$, where the solid line is the fit. (c) Collimation length as a function of D for different values of P.

Images of the top view are shown in Fig. 5(a) for a given value of D=14.5 mm, and for three values of the incident power P: 50, 70 and 110 mW, from top to bottom. The focal plane, indicated with dashed lines in each image, is displaced forward as the power increases, and CL grows in consequence.

Figure 5(b) illustrates the change in CL as a function of P for fixed values of D: 14.5, 15.5 and 16.5mm. Here we can see that, when P increases, the sample should be placed closer to MO in order to enhance the CL, which is consistent with Fig. 4(c). In other words, as the focal plane enters deeper into the sample, the balance between beam convergence and self-induced defocusing is optimized. Whereas for D=14.5mm the collimation length monotonically increases with power, for D=16.5mm there is a maximum of CL at P=80mW and then it decreases as P keeps increasing, since the self-defocusing becomes dominant.

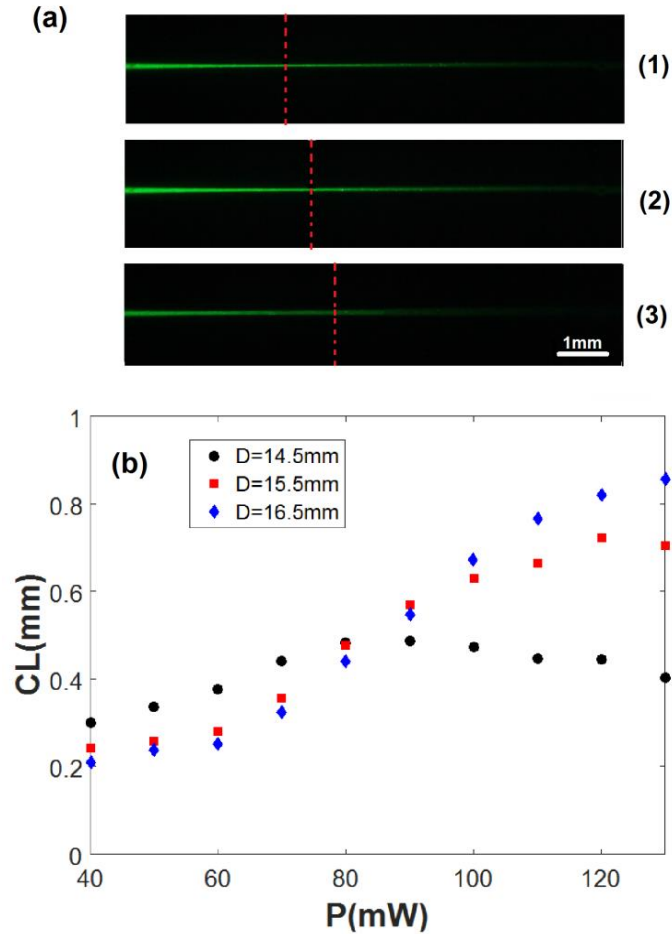

Figure 5. (a) Top view of the SCB obtained for a fixed value of D=14.5 mm, and different values of the incident power P: (1) 50, (2) 70 and (3) 110 mW. Dashed lines indicate the focal planes. In order to avoid saturation, an additional neutral density filter was used to obtain image (3). (b) Collimation length (CL) as a function of P for fixed values of the distance D.

On the other hand, we measured the response time (RT) of the sample for producing a stable SCB. For this purpose, we started with the beam initially blocked; then we unblocked it and observe the evolution of the transverse beam profile at the output face of the sample cell.

The whole process was recorded at a rate of 77frames per second, and we analyzed the video by doing a correlation between consecutive frames, as in the example illustrated in Fig. 6. While the beam is still blocked, only noisy fluctuations are obtained for the value of the correlation. Although the images appear black as the image A on the top, the correlation among them is not close to 1 because the camera itself is not blocked, and so it detects light leaks from the environment. At the moment when the beam is unblocked and impinges on the sample, it arises a minimum in the correlation (point B), due to the significant difference between frames (compare images A and B on the top). The transverse beam profile is then gradually modified due to the nonlinear effects (compare images B, C and D on the top), until reaching a steady state when a SCB is already formed (compare images D and E on the top), which appears as a plateau in Fig. 6. Point D in the plot represents the beginning of the plateau, when the correlation reaches its maximum value of ~1. The time difference between points B and D (indicated with the dashed lines) characterizes the RT, whose average value was found to be RT<0.07s±0.01s. It is worth to remark that, although the nonlinear absorption was found negligible, there is indeed a large linear absorption owed to the resonance of the gold nanoparticles at the laser wavelength (Fig. 1(b)), which is responsible of the thermal nonlinear effect, as explained before. In addition, there are scattering losses due to the presence of the particles in the medium. In consequence, only about 5 to 10% of the incident light reaches the output face of the sample cell.

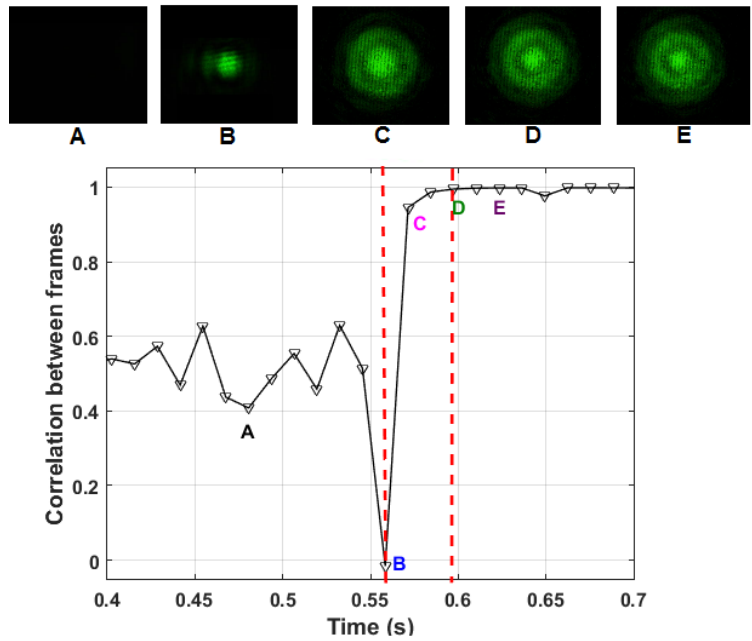

Fig. 6. One example of the curves used to estimate the response time of the sample by means of correlation between consecutive frames of a video showing the formation of SCB.

## 3. Control of a probe beam

Finally, we investigate an application of the SCB to control a weaker probe beam of a different wavelength ($\lambda_{probe}$=632.8 nm), which does not induce a nonlinear optical response in the sample by itself. A diagram of the experimental setup is shown in Fig. 3. The green laser acts as a pump beam ($\lambda_{pump}$ =532 nm) to induce a negative-lens effect in the medium. The propagation axes of both beams are set parallel by means of a beam combiner (BC), placed before the microscope objective (MO) that focuses both beams into the sample.

Images of the pump and probe beams are presented in the top and bottom rows of Fig. 7, respectively, for increasing values of the input power of the pump beam, corresponding, from left to right to: 1, 70, 90 and 110 mW. As we can see, both beams gradually shrink as the input power increases, while a set of concentric rings about the central spot starts to appear. The number and spacing of the rings vary with power, clearly differing from the naturally diffracting beams observed in the linear regime, when P=1mW. Ring patterns in the far field are typical when the optical nonlinearities have a thermal origin (26, 27). Noteworthy, the probe beam mimics the profile of the pump. The total transmitted power of the probe is always ~18%, independently of the incident power of the pump beam. The losses can be attributed to two main factors: scattering due to the particles and linear absorption, which is not negligible at this wavelength (see Fig. 1(b)).

It is worth to remark that the probe beam is not guided by a total internal reflection effect, since the refractive index has actually a minimum at the beam axis, but it is rather affected by the negative lens induced by the pump beam. Therefore, the focal plane of the probe is shifted forward along with that of the pump beam, except perhaps for a chromatic aberration. This also explains the high similitude between the transverse profiles of the probe and the pump.

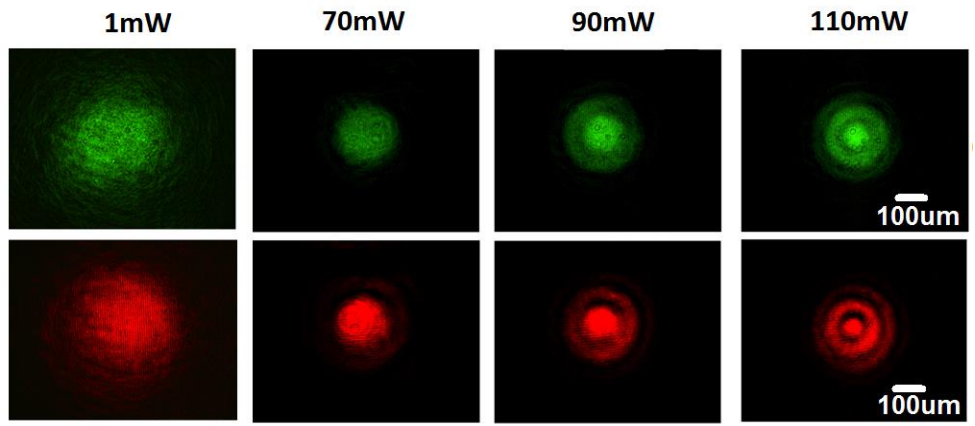

Figure 7. Output intensity profiles of the pump (top row) and probe (bottom row) beams for different pump powers indicated on the top.

Furthermore, the pump beam can be steered by shifting its propagation axis laterally, while keeping it parallel to the optical axis of MO. This is achieved by moving mirror M1, which is placed on a linear translation stage. When displaced, this beam passes through the same point at the focal plane, but its propagation axis is tilted with respect to the axis of the probe beam by a specific angle $\theta$. The total angular range attained in this way is approximately $-2.5° \leq \theta \leq 2.5°$. Once the pump beam is deviated, as illustrated on the left column of Fig. 8(a), the probe beam follows it, as can be seen on the right column of Fig. 8(a) for the top view, and in Fig. 8(b), corresponding to the output transverse profile, for a pump power of 110mW. In this case, the tilt of the pump plays the role of a misalignment of the self-induced lens with respect to probe axis, giving rise to a deflection of the latter following the former. Nonetheless, the misalignment introduces a comma aberration to the probe, as it is apparent from the images on the right of Fig. 8(b), which establishes a limit to the range of angular deviation.

Therefore, the pump beam not only controls the collimation length of the probe beam, but it is also able to steer it. The probe beam can be steered within a range of about 3° before becoming spoiled owed to a large aberration.

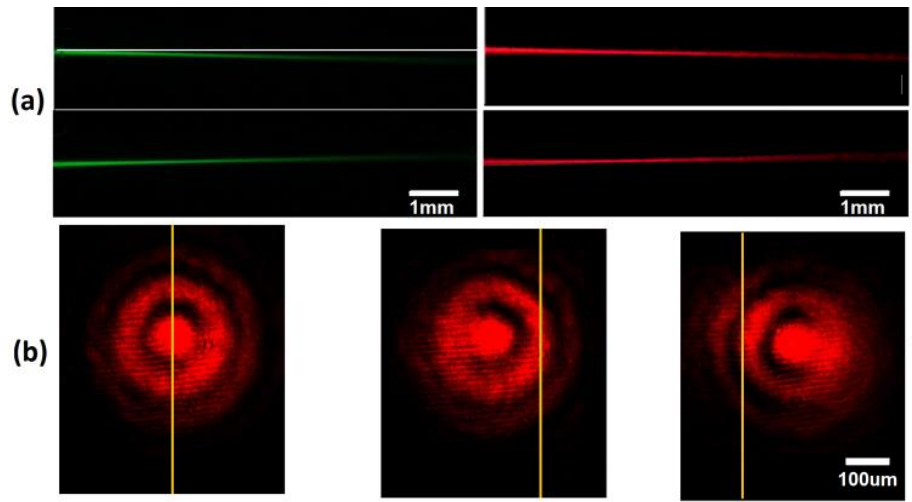

Figure 8. (a) Steering of the pump beam (left column) and the consequent deviation of the probe beam along the tilted path of the pump (right column); both beams propagate at angles of deflection of −1.25° (top), and 1.25° (bottom). (b) Transverse beam profile of the probe beam at the output face of the sample cell when the pump beam is steered. The yellow line shows the reference position when there is no tilt (0°). A tilt of -1.25° (left) and 1.25° (right) is achieved by steering the pump.

## 4. Conclusions

Gold nanoparticles have been successfully synthesized using a green methodology that is simple and effective. In this study, *Bacopa procumbens* extract and three compounds families obtained from it (glycosides, coumarins and saponins) have been used as a reducing agent for the synthesis of four colloidal suspensions of gold nanoparticles. We obtained gold nanoparticles with nearly spherical shape, with average diameters between 3 and 5 nm for the different samples. The biosynthesized colloidal suspensions of gold nanoparticles are fairly stable; one of them remained without visible changes for up to three months.

All the samples exhibit a negative nonlinear optical response of thermal origin, which is consistent with previous observations (19-21). This is different from the nonlinear optical response reported in Ref. (14), where the nonlinearity is mainly attributed to the redistribution of particles caused by the optical gradient force exerted by the illuminating beam. In our case the particles are so small that the optical forces are negligible and do not play a role.

On the other hand, we demonstrated that the nonlinear optical properties of the biosynthesized nanocolloids can be exploited to generate self-collimated beams. These are beams whose collimation length is about six times larger than that obtained in the linear regime, for instance, with a relative low input power (between 70 and 120mW). Self-collimated beams are obviously different from bright spatial solitons, since the self-defocusing medium rather produces a negative-lens effect that shifts the focal plane forward, diminishing the beam divergence thereafter. Importantly, the response time of the gold nanocolloid in terms of SCB formation is less than 0.07s, which is very fast for photonic applications compared with other nonlinear materials like photorefractive and nematic liquid crystals (28,29).

Furthermore, SCBs can be effectively used to control the path of a weaker probe beam of a different wavelength. This is attributed to the fact that the probe is affected by the change in the refractive index of the medium induced by the pump, and thus it also experiences a negative-lens effect. In consequence, the transverse intensity profile of the probe beam is dramatically modified as the power of the pump increases. While the probe mimics the profile of the pump, approximately 18% of its input power can be transmitted through the end of the sample cell. Moreover, we also demonstrated the capability of steering the probe beam over an angular range close to 3°, by controlling the incidence angle of the pump beam with an input power of 110mW. Less input power allows a lower steering range of the probe beam. Noteworthy, this is the first report, to the best of our knowledge, of control of a probe beam by means of self-induced thermal lens effect.

In summary, the biosynthesized colloidal suspensions of gold nanoparticles presented here exhibit a high nonlinear nonlocal optical response at relatively low input powers compared with dielectric colloids, while having a very fast response. These characteristics make this kind of media very suitable candidates for photonic applications.

**Acknowledgments** Authors acknowledge support from DGAPA-UNAM, grants PAPIIT IN114517, IA104917 and IPN with the Project SIP20170554. A. Balbuena-Ortega acknowledges support of CONACYT-Mexico